\documentclass[conference]{IEEEtran}
\IEEEoverridecommandlockouts
\usepackage[draft]{hyperref}
\usepackage{amsmath,amsfonts}
\usepackage{algorithmic}
\usepackage{graphicx}
\usepackage{textcomp}
\usepackage{xcolor}
\usepackage{fancyhdr}
\PassOptionsToPackage{hyphens}{url}
\usepackage{hyperref}
\usepackage{subfig}
\usepackage[normalem]{ulem}
\usepackage{subfloat}
\usepackage{pifont}
\usepackage{multirow}
\usepackage[ruled]{algorithm2e}
\usepackage{graphicx}
\usepackage{caption}
\usepackage{subfloat}
\usepackage{cite}
\usepackage{titlesec}

\def\BibTeX{{\rm B\kern-.05em{\sc i\kern-.025em b}\kern-.08em
    T\kern-.1667em\lower.7ex\hbox{E}\kern-.125emX}}
    
\begin{document}

\title{Brief Industry Paper: The Necessity of Adaptive Data Fusion in Infrastructure-Augmented Autonomous Driving System}

\vspace{-4mm}
\author{\IEEEauthorblockN{Shaoshan Liu\IEEEauthorrefmark{1}\IEEEauthorrefmark{2}, Jianda Wang\IEEEauthorrefmark{1}\IEEEauthorrefmark{3}, Zhendong Wang\IEEEauthorrefmark{3}, Bo Yu\IEEEauthorrefmark{2}, Wei Hu\IEEEauthorrefmark{2}, Yahui Liu\IEEEauthorrefmark{2},\\ Jie Tang\IEEEauthorrefmark{4}, Shuaiwen Leon Song\IEEEauthorrefmark{5}, Cong Liu\IEEEauthorrefmark{3}, Yang Hu\IEEEauthorrefmark{3}\IEEEauthorrefmark{6}}
\IEEEauthorblockA{\IEEEauthorrefmark{2}PerceptIn, USA\\
\IEEEauthorrefmark{3}The University of Texas at Dallas, USA\\
\IEEEauthorrefmark{4}South China University of Technology, China\\
\IEEEauthorrefmark{5}University of Sydney, Australia\\
\IEEEauthorrefmark{6}Corresponding author: Yang.Hu4@utdallas.edu
}
}
\vspace{-4mm}

\maketitle

\renewcommand{\thefootnote}{\fnsymbol{footnote}}
\footnotetext[1]{Authors with equal contribution}

\begin{abstract}
This paper is the first to provide a thorough system design overview along with the fusion methods selection criteria of a real-world cooperative autonomous driving system, named \textit{Infrastructure-Augmented Autonomous Driving} or IAAD. We present an in-depth introduction of the IAAD hardware and software on both road-side and vehicle-side computing/communication platforms. We extensively characterize the IAAD system in the context of real-world deployment scenarios and observe that the network condition fluctuates along the road is currently the main technical roadblock for cooperative autonomous driving. To address this challenge, we propose new fusion methods, dubbed "inter-frame fusion" and "planning fusion" to complement the current state-of-the-art "intra-frame fusion". We demonstrate that each fusion method has its own benefit and constraint. Adaptively choosing the fusion method according to the real-world condition will benefit the SoV without the violation of the SoV’s safety requirements.
\end{abstract}

\vspace{-3mm}
\section{Introduction}
The cooperative autonomous driving approach benefits from the collaboration between intelligent roads and smart vehicles enabled by emerging wireless communication technologies \cite{liu2021towards,liu2019edge}. It is not only more reliant but also more resilient in driving decision making compared to the traditional on-vehicle-only autonomous driving approaches, thanks to the driving-augmented information transmitted from the roadside systems. This paper provides the first thorough deployment experiences of an autonomous driving system enabled by intelligent roadside assistance, called \textit{Infrastructure-Augmented Autonomous Driving (IAAD) system}. 

The IAAD infrastructure consists of the roadside-deployed perception augmentation system, named \textit{System-on-Road (SoR)}, and the on-vehicle driving automation system, named \textit{System-on-Vehicle (SoV)}. The SoR aims to extend SoV’s "vision" by transmitting the perception information to the SoV and it equips with the upgraded counterparts. By leveraging the integrated information from the SoR and the SoV, the IAAD system can make safer and more efficient driving decisions in real time than the on-vehicle-only autonomous system.

Although the current IAAD design is promising, it is also very complex and comes with numerous real-world challenges. Our characterization study paints a frustrating picture that even the state-of-the-art version of the IAAD system configuration \cite{chen2019f} is far below the standard for feasibility. 

\textbf{\textit{To the best of our knowledge, there is no existing work that focuses on the network latency and jitter affect on the performance of a real-deployed IAAD system.}} 
Since the SoV-SoR system is a cooperative system, a “synchronous process” is required for the SoV to wait for the data from the SoR before coming to the next pipeline module. The time when the SoR arrives at the SoV will be significantly affected by the network condition in the real-world. We observe that the real-world wireless network condition is complex and ever-changing. Specially, we find that the state-of-the-art fusion mechanism (intra-frame fusion) that fuses the SoR data to the SoV data with the same time stamp at the tracking module on the SoV cannot adequately handle the complex external factors including high network latency and jitter. Our deployment data indicates that the deadline miss ratio can be as high as 30\%. 

Based on our field experimental experience, we pinpoint that the system tolerance to the network impacts could be improved by carefully choosing the data fusion method. Therefore, to tackle the realtime network fluctuation challenge in IAAD design, we propose new fusion methods, dubbed "inter-frame fusion" and "planning fusion" to complement the current state-of-the-art "intra-frame fusion". 

We demonstrate that each fusion method has its own benefit and constraint.Adaptively choosing the fusion method according to the real-world condition will benefit the SoV without the violation of the SoV’s safety requirements. 

\section{Real-world IAAD Testbed} \label{Setup}
The infrastructure-augmented autonomous driving (IAAD) system exhibits promising enhancement for the safety of the autonomous driving. \textbf{\textit{To the best of our knowledge, this is the first work that presents a real-world IAAD system and the corresponding thorough performance analysis.}} 

In this section, we first describe the IAAD’s system hardware configuration and then introduce its algorithmic components and software pipeline in an end-to-end manner. 

\subsection{System Hardware Deployment}

\newcommand{\tabincell}[2]{\begin{tabular}{@{}#1@{}}#2\end{tabular}}
\begin{table}[t]\small
\centering
\caption{Hardware setup of IAAD.}
\vspace{-2mm}
\scalebox{0.80}{
\begin{tabular}{|l|c|c|}
\hline
  & SoV & SoR\\\hline
\tabincell{l}{Industrial PC}&   \tabincell{l}{CPU: Intel i9-9900K \\GPU: GTX 1060 x1 \\ Memory: 32GB} & \tabincell{l}{CPU: Intel i9-9900K \\GPU: GTX 1080Ti x1\\ Memory: 32GB} \\\hline
\tabincell{l}{Sensors}&  \tabincell{l}{64-line LiDAR, \\camera x2, and Radar, \\GPS/IMU} &  \tabincell{l}{Lightweight: 80-line LiDAR\\Heavyweight: 80-line LiDAR, \\camera x2, and Radars} \\\hline
\tabincell{l}{Communication} & OBU & RSU \\\hline
\end{tabular}}
\vspace{-4mm}
\label{tb:iaad}
\end{table}

\begin{figure*}[t]
\vspace{-4mm}
\centering
            \includegraphics[width=0.80\linewidth]{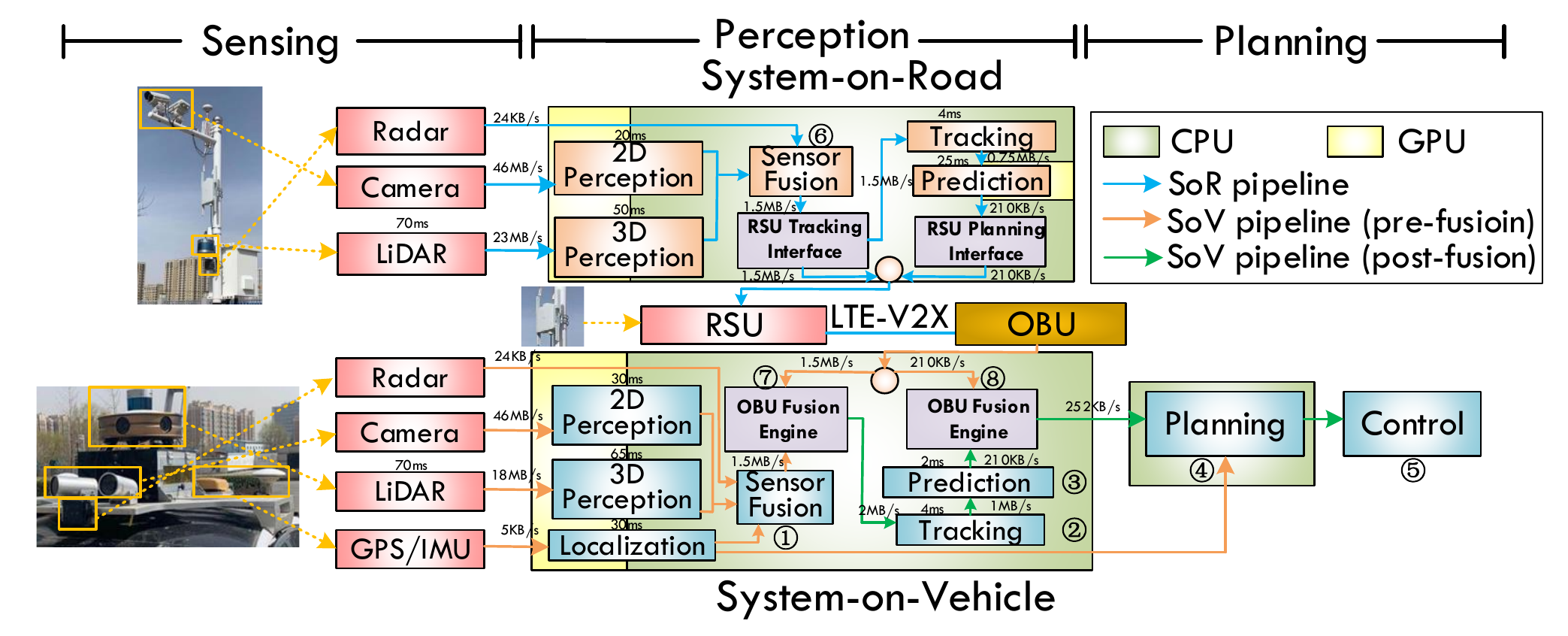}
            \vspace{-4mm}
            \caption{Overview of our infrastructure-augmented autonomous driving system.}\label{Systemview}
            \vspace{-6mm}
\end{figure*} \normalsize

\noindent\textbf{Hardware Setup:} Table \ref{tb:iaad} lists the essential hardware configuration of our IAAD system. Our IAAD system consists of an on-vehicle driving automation system (System-on-Vehicle, SoV) and roadside-deployed driving augmentation system (System-on-Road, SoR). The SoV consists of an Industrial PC (IPC), a series of sensors, a communication component (on-board unit, OBU) as well as the ECU and the CAN bus connecting all components. ECU and CAN bus are standard autonomous components. The SoR consists of consists of an Industrial PC (IPC), a series of sensors, a communication component (roadside Unit, RSU). Different from the SoV, the SoR's IPC is configured with a high-end Nvidia 1080Ti GPU and two optional sets of perception sensors. The SoR can be configured as heavy-weight SoR and light-weight SoR for the different deployment purposes. The heavy-weight SoR equips with cameras and LiDARs while the light-weight SoR only equips with LiDAR. This is because the LiDAR provides more accurate perception compared to the camera, especially during the night time. The RSUs and OBUs communicate directly to facilitate the fusion of the SoR data and the SoV data on the vehicle-side. The communication protocol is based on LTE-V2X currently and will support 5G-V2X in the future. 

\subsection{IAAD, An End-to-end Description}
Fig. \ref{Systemview} presents an end-to-end illustration of our IAAD system. The main components' latency and throughputs are demonstrated as well. Overall, both systems run a customized kinetic ROS middleware and a series of algorithmic modules. Specifically, our full-fledged L4 SoV includes Continental ARS 408-21 mmWave radars for detecting and tracking objects, cameras with 1920x1080 pixels for sensing and 2D perception module (based on YOLOv5), 64-line LiDAR and 3D perception module (based on PointPillars), GPS/IMU and localization module (based on NDT). The 2D perception and 3D perception results, as well as the outputs from mmWave radars, are combined to generate a comprehensive perception view in SoV-side sensor fusion module \ding{192}. The fused perception results are subsequently fed into the tracking module to keep track of each detected object \ding{193}. The purpose of tracking is to establish a temporal relation of the same object across multiple time frames so that the SoV gets to know the trajectory of the target object. Then the tracking results are fed to the prediction module, whose purpose is to predict the future trajectory of the target object based on its past trajectory \ding{194}. Finally, the prediction results are fed to the planning module to make behavioral and motion planning decisions \ding{195}. After a planned trajectory is generated for SoV, the control module will work with the vehicle chassis to execute the trajectory \ding{196}.

In the meantime, the SoR detects and tracks objects with Continental ARS 408-21 mmWave radars. The SoR also performs 2D perception for the camera data and 3D perception for the LiDAR data. The SoR sensor fusion then generates the perception outputs \ding{197}. The perception outputs will be processed by tracking module and prediction module. Note that the SoR performs a heavy-weight prediction (H-prediction) \cite{heavypre}. Upon the various requests from SoV, either the perception outputs or prediction outputs of SoR could be transmitted to the SoV through LTE-V2X. These outputs can be fused with the SoV data and then fed to the corresponding modules on the SoV (e.g., either tracking \ding{198} or planning \ding{199}).
\begin{figure}[t]
\vspace{-2mm}
\subfloat[Latency and jitter]{
\centering
            \includegraphics[width=4.0cm, height = 2.8cm, trim=2 0 2 0, clip]{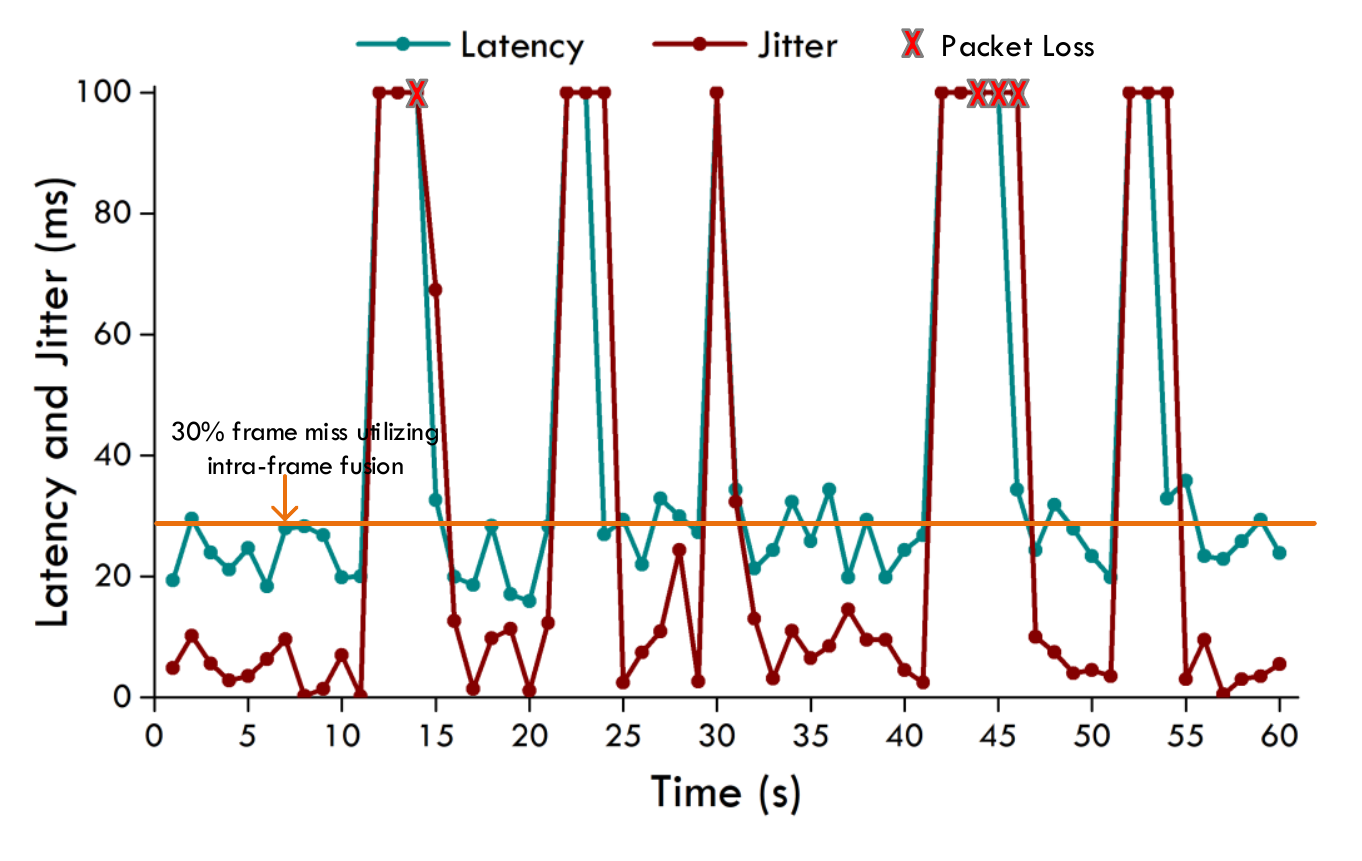}            
            \label{Latencyl}}
\subfloat[Light and heavy prediction performance comparison]{
\centering
            \includegraphics[width=4.0cm, height = 2.8cm, trim=2 0 2 0, clip]{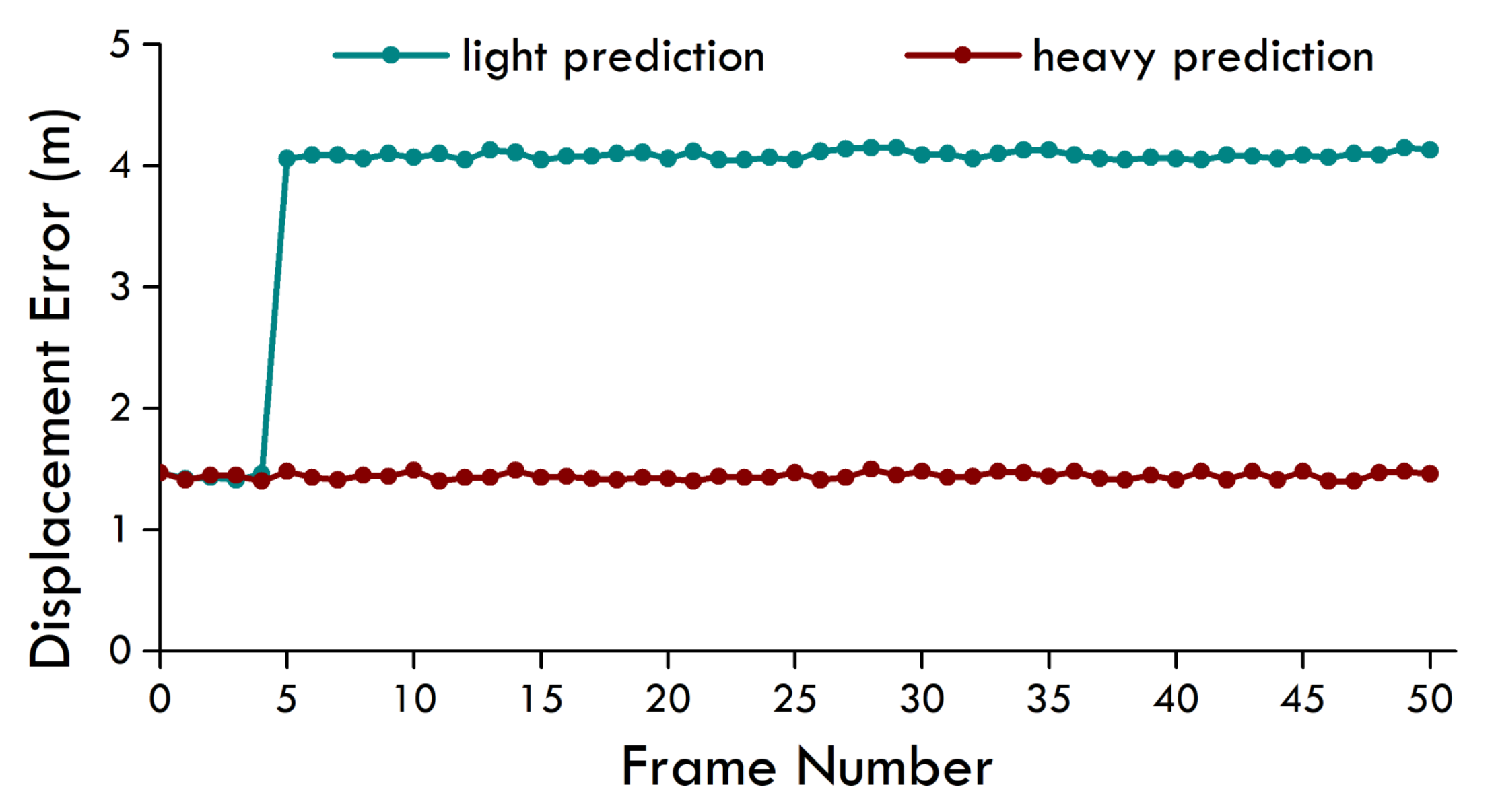}
             \label{displacementerror}
}
\vspace{-1mm}
\caption{Latency and jitter of real-world network and Comparison of light and heavy prediction}
\vspace{-3mm}
\end{figure} \normalsize

\vspace{-3mm}
\section{The Real-world Network fluctuation challenge} \label{characterization}
Our deployment experiences demonstrate that it is challenging to enjoy the benefit of long-range detection capability and high prediction accuracy in an IAAD system due to the real-world network fluctuation constraint. In this section, we thoroughly explore the  challenges of an IAAD system provoked by the real-world network system constraint.

\subsection{The New Metric Reflecting the Prediction Accuracy}
To demonstrate the IAAD's safety quantitatively, we firstly define a metric \textbf{\textit{displacement error}}, which indicates the deviation between the predicted driving path and the ground-truth (i.e. ideal) driving path. The vehicle's safety will be high when the displacement error keeps low. The equation \ref{equ0} illustrates the definition of the displacement error:

\begin{footnotesize}
\vspace{-3mm}
\begin{equation} \label{equ0}
Displacement \  error = \frac{\sum\limits_{i=1}^{N} \lvert Location(i)_{prediction}-Location(i)_{ideal}\lvert}{N}
\end{equation}
\vspace{-4mm}
\end{footnotesize}

\noindent where the $Location(i)_{prediction}$ is the predicted location of the object i, the $Location(i)_{ideal}$ is the ground truth location of the object i and N is the total number of the detected objects. Our Field deployment results are illustrated in Table \ref{tb:accoftwofusions}. In this case, we assume every frame can arrive "in time" without the effect of the network fluctuation. We can observe that the displacement error decreases as the number of input frames increases. The displacement error will stay stable when the input frames achieve 20 frames. Consequently, our system will maintain a 20 frames window for the path prediction and will update the path prediction result when each new frame flows in. 

\begin{table}[b]\small
\vspace{-3mm}
\centering
\small
\caption{Field deployment results}
\vspace{-2mm}
\scalebox{0.8}{
\begin{tabular}{|l|l|l|l|l|}
\hline
\multicolumn{1}{|c|}{\multirow{2}{*}{\begin{tabular}[c]{@{}c@{}}Input Frames \\ Number\end{tabular}}} & \multicolumn{1}{c|}{\multirow{2}{*}{\begin{tabular}[c]{@{}c@{}}Displacement \\ Error (meter)\end{tabular}}} & \multicolumn{3}{c|}{Distance (meter) between vehicle and object} \\ \cline{3-5} 
\multicolumn{1}{|c|}{}                                                                                & \multicolumn{1}{c|}{}                                                                                       & Without SoR           & Light SoR           & Heavy SoR          \\ \hline
5                                                                                                     & 4.333                                                                                                       & 63                    & 163                  & 313                \\ \hline
10                                                                                                    & 3.053                                                                                                       & 56                    & 156                  & 306                \\ \hline
\textcolor{red}{15}                                                                                                    & \textcolor{red}{2.192}                                                                                                       & 49                    & 149                  & 299                \\ \hline
\textcolor{red}{20}                                                                                                    & \textcolor{red}{1.360}                                                                                                        & \textcolor{red}{42}                    & \textcolor{red}{142}                  & \textcolor{red}{292}                \\ \hline
25                                                                                                    & 1.358                                                                                                        & 35                    & 135                 & 285                \\ \hline
\end{tabular}}\label{tb:accoftwofusions}
\vspace{-5mm}
\end{table}

\subsection{Real-World Network Fluctuation Challenge to the state-of-the-art Fusion Method}
\label{challenge}

\noindent\textbf{Intra-frame Tracking Fusion Constraint:} The existing efforts on cooperative perception consider the tracking fusion as the state-of-the-art solution\cite{chen2019f}. In the tracking fusion method, the RSU in SoR will extract the sensor fusion module's output features and transmit them to SoV's OBU using an LTE-V2X connection. After receiving the data, the OBU will fuse the data with its own sensor fusion module's output and subsequently publish the fused data into the car-side software pipeline's tracking module. Under this situation, the SoV output will be fused with the SoR frame which has the same timestamp. We name this tracking fusion method as "intra-frame tracking fusion". Since the SoV-SoR system is a cooperative system, a “synchronous process” is required for the SoV to wait for the data from the SoR before coming to the next pipeline module. This synchronous process will require an extra time window on the SoV side to wait for the data from the SoR side. \textbf{\textit{The presumption of the existing work is that the frame from the SoR will always arrive the SoV in time. However, based on our field test, there are a bunch of cases when the SoV cannot receive the frame of the SoR in time due to the SoR latency caused by the network fluctuation.}} Since the autonomous driving exists the End-to-End (E2E) pipeline boundary and the output interval boundary, the maximum time window has to comply with the above two boundaries to guarantee the vehicle's safety. In the IAAD system, SoR latency caused by the network fluctuation will extend the E2E pipeline time and the output interval time. To avoid the violation of the two above boundaries, when the SoV cannot receive the SoR information within the time window, it needs to go through to the subsequent modules without the fusion to guarantee the safety of the vehicle. The E2E pipeline boundary and output interval boundary will be discussed as follows.

The E2E pipeline is defined as the duration from the time when the vehicle receives the sensing data to the time when the vehicle finishes its planning decision. Output interval is defined as the time between the two consecutive planning decisions. For the vehicle’s safety consideration, the E2E software pipeline latency and output interval both need upper boundaries. Thus, if the SoV cannot receive the data from the SoR "in time", the data from SoV has to skip the “data fusion” process to go through the pipeline individually in order to follow the requirement of the E2E pipeline and output interval boundary.

\noindent\textbf{Performance Degradation caused by the Real-world Network Fluctuation:} 
Fig. \ref{Latencyl} depict the network condition in our real test field. We can observe that the network latency fluctuates between 15 ms to 35 ms and sometimes will achieve 100ms. Even worse, in some cases the packets may totally get lost due to the bad network condition. Consequently, the SoR latency caused by the network fluctuation will probably lead to the violation of the E2E pipeline latency requirement. Besides, as shown in Fig. \ref{Latencyl}, we observe that the network jitter will significantly hurt the stability of the vehicle’s planning decision interval time, which will cause the violation of the vehicle's output interval boundary. Based on our field experience, the SoV will miss around 30\% of the data from the SoR when simply utilizing intra-frame tracking fusion method, as illustrated in the Fig. \ref{Latencyl}, which will cause the degradation of the performance of the IAAD system. 

\vspace{-1mm}
\subsection{Alternative Fusion methods to tackle the network fluctuation challenges}
In the previous section, we demonstrate the network fluctuation challenge confronted by utilizing the intra-frame tracking fusion method in the real world. Due to the E2E boundary and output interval boundary, this fusion method's requirement is strict. Since the network will fluctuate in the real-world, a bunch of the frames will be missed from the SoR, which will cause the performance degradation for our IAAD system. To deal with the network challenge and guarantee the IAAD safety, we take the step to explore a flexible data fusion framework, dubbed as \textbf{\textit{inter-frame tracking fusion and planning fusion}}. In this section, we will illustrate the benefit and constraint of our inter-frame tracking fusion and planning fusion method compared to the intra-frame tracking fusion. \textbf{\textit{We demonstrate that the proposed fusion methods have their own benefit and constraint, adaptively choosing the fusion method according to the real-world condition will benefit the SoV without the violation of the SoV’s safety requirements.}}

\begin{figure}[t]
\vspace{-8mm}
\subfloat[Complementary view field case]{
\centering
            \includegraphics[width=4.0cm, height = 2.8cm, trim=2 0 2 2, clip]{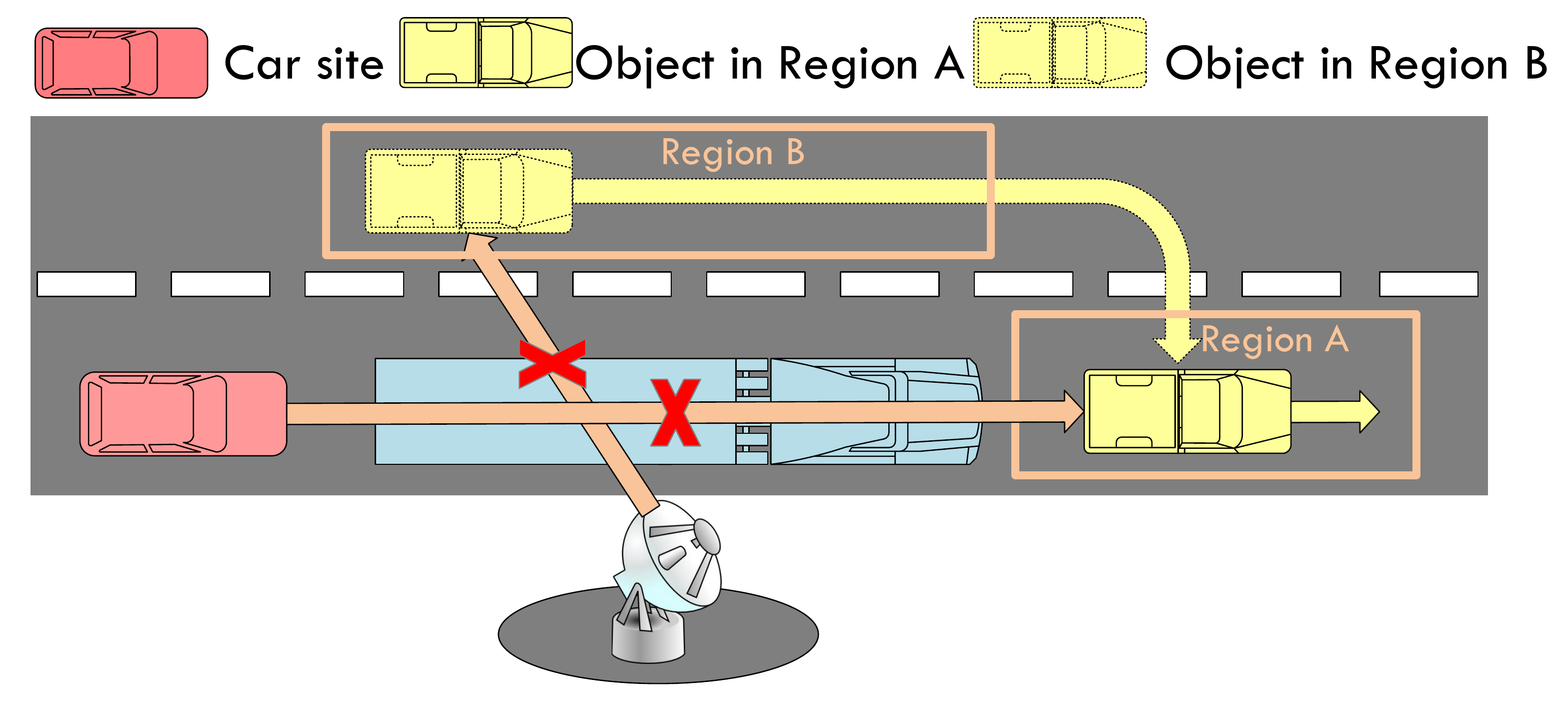}
            \label{Compleview}}
\subfloat[Accuracy comparision]{
\centering
            \includegraphics[width=4.0cm, height = 2.8cm, trim=2 0 2 2, clip]{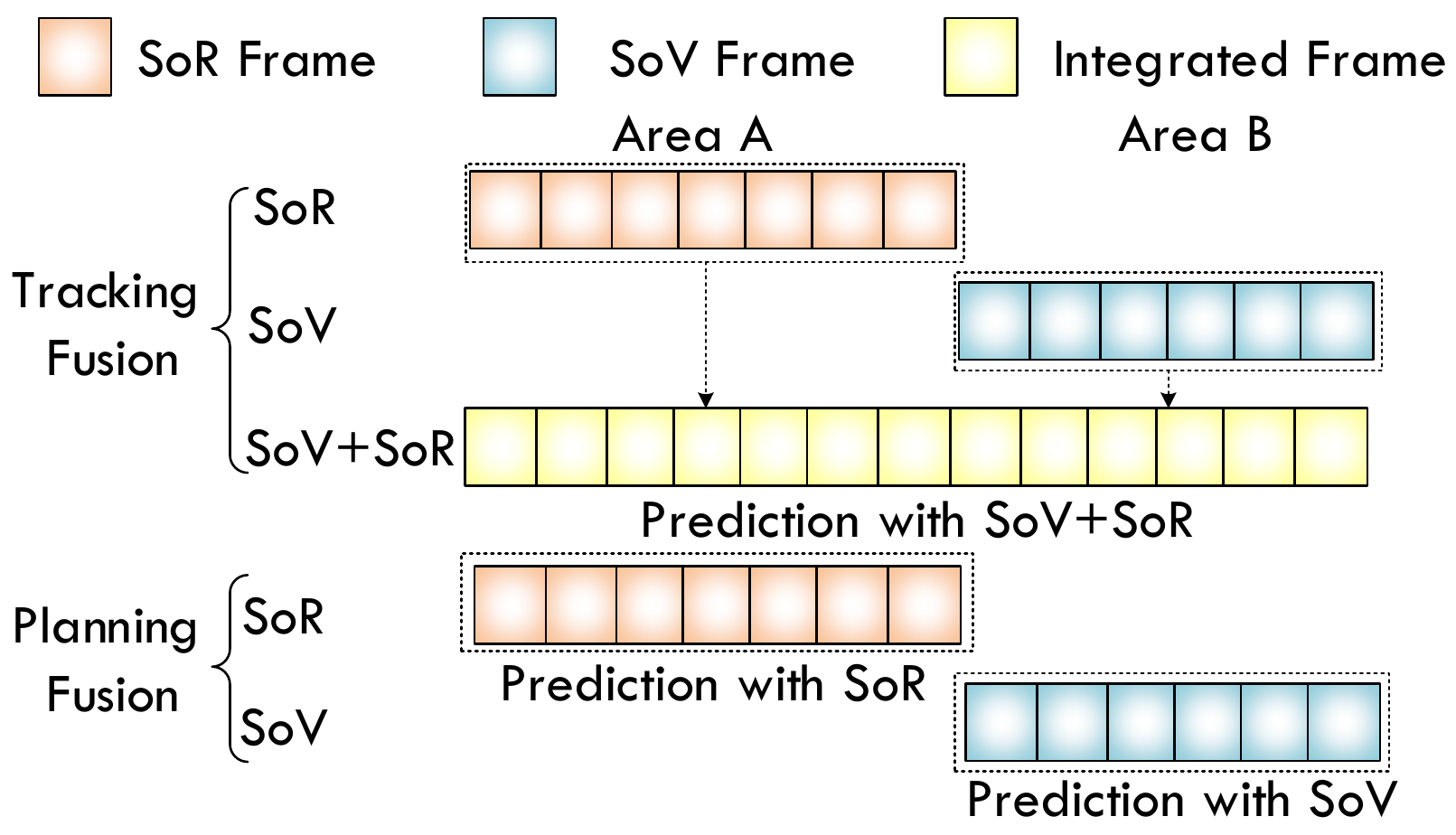}            
            \label{TracAccu}}
\vspace{-2mm}
\caption{Analysis for fusion effectiveness under different fusion position}
\vspace{-6mm}
\end{figure} \normalsize

\noindent\textbf{Benefits and Constraints of Inter-frame Tracking Fusion:} The intra-frame fusion incurs a strict waiting window for the SoV data. Based on our real deployment experience, the OBU in SoV can intermittently miss the waiting window due to some unavoidable network situations (e.g, base station handovers, congestion), which will cause the performance degradation in the system. To mitigate the affect, we explore the opportunity to utilize the previous SoR frame stored in the buffer to fuse with the current SoV frame if the intra-frame waiting window is missed. We name this mechanism as \textbf{\textit{inter-frame fusion}}.

Unavoidably, the prediction accuracy of the inter-frame fusion will decrease compared to the intra-frame fusion (the requirement is most strict while the accuracy is the highest). However, if the timestamp of the frame from SoR (stored in the buffer) is very close to the current timestamp, the prediction accuracy does not decrease significantly. Besides, the tolerance of the prediction accuracy is high when concerning the far-fledged objects detected by the SoR compared to the nearby objects detected by the SoV. Consequently, the inter-frame fusion will effectively decrease the performance degradation compared to the system without fusion under the situation where the current frame from the SoR is missed. 

However, the inter-frame fusion still has its constraint under the bad network condition, since the displacement error will become unacceptably high when missed frames accumulate due to the long-time network congestion. The tolerance time for the inter-frame fusion is around 0.6s, which will be adequate for the short time network congestion (i.e. base station handover) but still be not enough when long time network congestion occurs (i.e. go through tunnel region).

\noindent\textbf{Benefits and Constraints of Planning Fusion:} In our tracking fusion (intra-frame and inter-frame) mechanism, the post-perception data from the SoR will be combined with local data and processed by the tracking and \textit{local prediction module}. However, a poor network environment can cause discontinued provision of the SoR data due to the timeouts. The loss of augmentation of the SoR data will lead to a comparable displacement error in a much near distance with the object.

Since our local prediction module adopts a low-level prediction algorithm which only provides a prediction of next 0.5 seconds, the prediction accuracy will be significantly reduced if the network condition exacerbates, as shown in Fig. \ref{displacementerror}. We can observe that the displacement error of light prediction can maintain in a low-level only for the first 5 frames and then increases to a high-level in the following 45 frames. To solve this issue, our SoR equips with a heavy-weight prediction algorithm (\cite{heavypre}) which can provide trajectory prediction for the next 5 seconds. This provides an opportunity that the prediction results from the SoR with lower displacement error, could be transmitted to the SoV and be utilized by planning module of SoV in the next 5 seconds instead of 0.5 second. As shown in Fig. \ref{displacementerror}, the displacement error of heavy-prediction method keeps low in the entire 50 frames, and with a higher tolerance to the network latency. Motivated by this, we implement the planning fusion mechanism to overcome the performance degradation occurred in the tracking fusion method. By implementing the planning fusion, the tolerance time will be significantly extended compared to the tracking fusion.

Although the planning fusion will tolerate the congested network condition compared to the tracking fusion, the displacement error is higher than the tracking fusion. The reason for the higher displacement error is that the SoV's view and SoR's view are complementary with each other. Integrating SoR's data into a late stage in the SoV's software pipeline can cause the "spatial deficiency" for the system. For instance, as shown in Fig. \ref{Compleview}, when the object stays in region A, only the vehicle can detect the object (yellow vehicle) while the road cannot detect it. When the object goes into the region B, only the road can detect the object while the vehicle cannot detect it. As shown in Fig. \ref{TracAccu}, the tracking fusion can integrate the fragmentary data from both sides to do its prediction while planning fusion will have to rely on the SoV and the SoR to do the prediction with their own fragmentary data, which will lead to the degradation of the accuracy for the final planning output. Consequently, when the network condition becomes better, we need to switch the planning fusion back to tracking fusion to avoid the spatial deficiency of the planning fusion.

\vspace{-2mm}

\end{document}